\documentclass[prd,twocolumn]{revtex4}

\usepackage{graphicx, epsfig}
\usepackage{epstopdf}
\usepackage{color}
\usepackage{mathrsfs}
\usepackage{amsmath}
\usepackage{braket}

\newcommand{\be}{\begin{equation}}
\newcommand{\ee}{\end{equation}}
\newcommand{\bea}{\begin{eqnarray}}
\newcommand{\eea}{\end{eqnarray}}

\newcommand{\gapp}{\mathrel{\raise.3ex\hbox{$>$}\mkern-14mu
              \lower0.6ex\hbox{$\sim$}}}
\newcommand{\lapp}{\mathrel{\raise.3ex\hbox{$<$}\mkern-14mu
              \lower0.6ex\hbox{$\sim$}}}

\begin{document}
\title{Hawking-like radiation and the density matrix for an infalling observer during gravitational collapse }
\author{Anshul Saini, Dejan Stojkovic}
\affiliation{ HEPCOS, Department of Physics, SUNY at Buffalo, Buffalo, NY 14260-1500, USA}


\begin{abstract}
We study time-dependant Hawking-like radiation as seen by an infalling observer
during gravitational collapse of a thin shell. We calculate the occupation number of particles whose frequencies are measured in the
proper time of an infalling observer in Eddington-Finkelstein coordinates. We solve the equations for the whole process from the beginning of the collapse till the moment when the collapsing shell reaches zero radius. The radiation distribution is not thermal in the whole frequency regime, but it is approximately thermal for the wavelengths of the order of the Schwarzschild radius of the collapsing shell. After the Schwarzschild radius is crossed, the temperature increases without limits as the singularity is approached. We also calculate the density matrix associated with this radiation.
It turns out that the off-diagonal correlation terms to the diagonal Hawking's leading order terms are very important.
 While the trace of the diagonal (Hawking's) density matrix squared decreases during the evolution,
the trace of the total density matrix squared remains unity at all times and all frequencies.

\end{abstract}


\maketitle

\section{Introduction}
Hawking radiation from a black hole is an extensively studied subject. Yet, there is still a lot to learn about it. Most of the studies of Hawking radiation are done from the point of view of an outside observer who is watching outgoing radiation from a black hole. A very small fraction of the literature (see e.g. 
\cite{Greenwood:2008zg,Barbado:2016nfy,Barbado:2012pt,Barbado:2011dx,Kim:2009ha,Kim:2013caa}) is dedicated to the question of what an infalling observer would see. Naively, one might argue that an infalling observer should not register any Hawking radiation because of the equivalence principle since he is in a free fall state. However, we know that the equivalence principle is valid locally, where the locality is defined by curvature radius.  Therefore, while an infalling observer would not see all the particles which are excited in this space-time, he will definitely observe modes whose wavelengths are comparable or larger than his local curvature radius. Only those modes whose wavelength is shorter than the local curvature radius around him would remain invisible. It is therefore expected that an infalling observer would also register radiation.

In this paper we study radiation which is excited by a collapsing shell as observed by an infalling observer in Eddington-Finkelstein space-time foliation. Since the spacetime metric is time dependent during the collapse, modes of radiation of the field which propagates in such a background will be excited.
We will use the functional Schrodinger formalism to calculate the modes which are excited by the collapsing shell from the start of the collapse when the shell is larger than its own Schwarzschild radius, until the moment when the shell collapses to a point and singularity is formed. Unlike some of the previous work, we will not use any approximations and our results will be exact (although the final spectra are calculated numerically).  Since the functional Schrodinger formalism gives the wavefunction for the state of excited radiation, we will also calculate the relevant density matrix. If one keeps just the diagonal terms in the density matrix (i.e. particles) then the evolution does not appear to be unitary. However, if one includes the correlations between the diagonal modes then one recovers unitary evolution.

\section{The setup}
\label{metric}
We consider an infinitely thin collapsing shell of the total mass $M$ (we will see below that the total mass includes the rest mass, kinetic energy, and gravitational energy of self-interaction). For our purpose, the shell has constant energy per unit area, i.e. $\sigma =$const.  The collapsing shell induces a time-dependent metric. This time-dependent metric will in turn excite modes of radiation. We will study this radiation
in the space-time foliation of an infalling observer in the Eddington-Finkelstein coordinate system, which is not singular at the horizon. The metric outside of the collapsing shell is given by
\be
  ds^2=-\left(1-\frac{R_s}{r}\right)dv^2+2dvdr+r^2d\Omega^2, \hspace{2mm} r>R(v).
  \label{out_Met}
\ee
where $R(v)$ is the time-dependent radius of the shell, while $R_s = 2GM$. The infalling time parameter $v$ for the  Eddington-Finkelstein coordinate system is defined as
\be
 v=t+r^*
\ee
Here $r^*$ is the standard tortoise coordinate.
The interior of the shell is just a flat spacetime due to the Birkoff theorem. The interior metric is therefore Minkowski, though with the different time parameter, $T$,
\be
{ds}^2 =  - dT^{2}  +  dr^2 + r^2 d{\Omega}^{2} .
\label{in_Met}
\ee
The metric exactly at the shell can be written as
\be
ds^2 = - {d \tau}^2 + r^2 d{\Omega}^2
\label{at_Met}
\ee
where $\tau$ is the proper time of an observer who is sitting on the shell.
Using Eq.(\ref{out_Met}), Eq.(\ref{in_Met}) and Eq.(\ref{at_Met}), we can find the relations connecting time coordinates $T$ and $v$ with $\tau$  as
\be
  \frac{dT}{d\tau}=\sqrt{1+\left(\frac{dR}{d\tau}\right)^2}
  \label{dTdtau}
\ee
and
\be
  \frac{dv}{d\tau}=\frac{1}{B}\left(\frac{dR}{d\tau}-\sqrt{B+\left(\frac{dR}{d\tau}\right)^2}\right)
  \label{dvdtau}
\ee
where
\be
  B\equiv 1-\frac{R_s}{R}.
\ee
Eq.(\ref{dTdtau}) and Eq.(\ref{dvdtau}) can be used to relate $v$ and $T$ as
\be
\frac{dv}{dT} = \frac{1}{\sqrt{R_v^2 - 2 R_v +B}}
\label{vTrelation}
\ee
where $R_v = \frac{dR}{dv}$. We can further use Eq.(\ref{dvdtau}) to arrive at the relation
\be
R_v = \frac{B}{1-\sqrt{1+ \frac{B}{R_\tau^2}}}
\label{rvdynamics}
\ee
In \cite{Ipser:1983db}, it was shown that the total mass of the collapsing shell, $M$, is a conserved quantity of the system and it is given by expression
\be \label{M}
M =  4 \pi \sigma R^2 \left( \sqrt{1+ {R_{\tau}}^2} - 2\pi G \sigma R\right) .
\ee
Obviously, the total mass includes the rest mass $M_{\rm rest}=4 \pi \sigma R^2$, kinetic energy represented by $R_{\tau}^2$, and gravitational energy of self-interaction
$GM_{\rm rest}^2/(2R)$.
The conservation of $M$ during the evolution can simply be checked by taking the time derivative and using the equations of motion.
We can use Eq.~(\ref{M}) to express $R_{\tau}$ as
\be
R_{\tau} = \sqrt{{\left( \frac{M}{4 \pi G R^2}+ 2 \pi \sigma G R\right)}^2-1} .
\label{rconserve}
\ee
Eq.~(\ref{rconserve}) together with Eq.~(\ref{rvdynamics}) can be used to find the evolution of the shell as a function of time. Once we know the classical dynamics of collapsing shell, we can proceed further by putting  a scalar field in this time dependent background and look for its excitations. The action for such configuration  is
\be
  S=\int d^4x\sqrt{-g}\frac{1}{2}g^{\mu\nu}\partial_{\mu}\Phi\partial_{\nu}\Phi .
  \label{action}
\ee
We can expand the scalar field in terms of  real basis functions
\be
  \Phi=\sum_k a_k(v)f_k(r) .
  \label{mode_ex}
\ee
The exact form of the basis functions, $f_k(r)$, will not matter, and the coefficients $a_k$ will play the role of the dynamical variables after quantization in the functional Schrodinger formalism.
Since the metric inside and outside of the shell have different forms, the action should be written separately for each case. The action for interior of the shell is
\be
  S_{in}=2\pi \int dT\int_0^{R(v)}drr^2\left[-(\partial_T\Phi)^2+(\partial_r\Phi)^2\right] .
  \label{S_in}
\ee
Similarly, the action outside of the shell is
\begin{align}
  S_{out}=2\pi\int dv\int_{R(v)}^{\infty}drr^2&\Big{[}\partial_v\Phi\partial_r\Phi+\partial_r\Phi\partial_v\Phi\nonumber\\
  &+\left(1-\frac{R_s}{r}\right)(\partial_r\Phi)^2\Big{]}.\label{S_out}
\end{align}
Using Eq.(\ref{vTrelation}), we can rewrite $S_{in}$ in term of coordinate $v$. This yields
\begin{align}
  S_{in}=2\pi\int dv\int_0^{R(v)}drr^2&\Big{[}-\frac{1}{\sqrt{{R_v}^2 - 2 R_v +B}}(\partial_v\Phi)^2\nonumber\\
  &+\sqrt{{R_v}^2 - 2 R_v +B}(\partial_r\Phi)^2\Big{]} . \label{S_in_v}
\end{align}
The total action for the scalar field is now
\be
S = S_{in} + S_{out} .
\ee
After substituting the expressions for $S_{in}$ and $S_{out}$ we have
\begin{align}
  S =&2\pi\int dv\Big{[}\int_0^{R}drr^2\Big{(}-\frac{1}{\sqrt{{R_v}^2 - 2 R_v +B}}(\partial_v\Phi)^2\\
  &+\sqrt{{R_v}^2 - 2 R_v +B}(\partial_r\Phi)^2\Big{)}\nonumber\\
  &+\int_{R}^{\infty}drr^2\partial_v\Phi\partial_r\Phi+\int_{R}^{\infty}drr^2\partial_r\Phi\partial_v\Phi\nonumber\\
  &+\int_{R}^{\infty}drr^2\left(1-\frac{R_s}{r}\right)(\partial_r\Phi)^2\Big{]} .
\end{align}
This action can be rewritten in terms of the functions $a_k$ with the help of Eq.~(\ref{mode_ex}) as
\begin{align}
  S = \int dv&\Big{[}\frac{1}{2}\frac{1}{\sqrt{{R_v}^2 - 2 R_v +B}}\dot{a}_k{\bf A}_{kk'}\dot{a}_{k'}\nonumber\\
  &-\frac{1}{2}\sqrt{{R_v}^2 - 2 R_v +B} \ {a}_k{\bf D}_{kk'}{a}_{k'}-\frac{1}{2}\dot{a}_k{\bf Y}_{kk'}a_{k'}\nonumber\\
  &-\frac{1}{2}a_k{\bf Y}_{kk'}^{-1}\dot{a}_{k'}-\frac{1}{2}a_k{\bf C}_{kk'}a_{k'} \Big{]} ,
\end{align}
where $\dot{a}=da/dv$, while ${\bf A}$, ${\bf Y}$, ${\bf D}$ and ${\bf C}$ are matrices given by
\begin{align}
  {\bf A}_{kk'}=-4\pi\int_0^{R}drr^2f_k(r)f_{k'}(r),\\
  {\bf D}_{kk'}=-4\pi\int_0^{R}drr^2f'_k(r)f'_{k'}(r).\\
  {\bf Y}_{kk'}=-4\pi\int_{R}^{\infty}drr^2f_k(r)f'_{k'}(r),\\
  {\bf C}_{kk'}=-4\pi\int_{R}^{\infty}drr^2\left(1-\frac{R_s}{r}\right)f'_k(r)f'_{k'}(r).
\end{align}
The explicit form of these matrices depend on the functional form of the basis $f_k(r)$. However, we are interested in the dynamics of excitation of the modes not in their internal structure, so we will not need their explicit forms. It can be seen from above relations that matrices are symmetric and real, and in particular ${\bf Y}={\bf Y}^{-1}$. Let's define a shorthand
\be \label{lambda}
\lambda(v) \equiv \sqrt{{R_v}^2 - 2 R_v +B}
\ee
where $R_v =\frac{dR(v)}{dv}$. We can rewrite the action as
\begin{align}
  S = \int dv&\Big{[}\frac{1}{2 \lambda}\dot{a}_k{\bf A}_{kk'}\dot{a}_{k'} - \frac{\lambda}{2}{a}_k{\bf D}_{kk'}{a}_{k'} \nonumber\\
  &-\frac{1}{2}{\bf Y}_{kk'}\left(\dot{a}_ka_{k'}+a_k\dot{a}_{k'}\right)-\frac{1}{2}a_k{\bf C}_{kk'}a_{k'}\Big{]}.\label{Action}
\end{align}
Assuming that all matrices can be diagonalized simultaneously (which we admit is a strong assumption) we get
\begin{equation}
\mathcal{L} = \frac{1}{2 \lambda} m {\dot{b}}^2 - y b \dot{b} - \frac{1}{2}(\lambda l + k)b^2
\end{equation}
where $m$, $y$, $l$ and $k$ denote eigenvalues of ${\bf A}$, ${\bf Y}$, ${\bf D}$ and ${\bf C}$ respectively, whereas $b$ is the eigenmode (which is a linear combination of the original modes $a$). While there are infinitely many original modes $a_k$, they all have the same form of Lagrangian after diagonalization. Thus, it will be enough to solve the dynamics for one eigenmode $b$. Using standard classical calculation the Hamiltonian of the system which follows from this Lagrangian is
\begin{equation}
H = \frac{\lambda}{m}\left[ \frac{\pi^2}{2} + \pi y b + \frac{1}{2} (y^2 + l m +\frac{k m }{\lambda})b^2 \right] ,
\end{equation}
where $\pi$ is the generalized momentum corresponding to the dynamical variable $b$.
It is important to mention that we did not use any approximation, so this is full Hamiltonian for the scalar field propagating in the background of a collapsing shell w.r.t the infalling observer. In the next section, we quantize this Hamiltonian using standard quantization technique to obtain the wavefuntion which contains all the information about the excited modes, i.e. radiation.

\section{Quantization}

The first step in quantization is to promote the momentum $\pi$ into an operator, i.e. $\hat{\pi} = - i \frac{\partial}{\partial b}$. Next, we have to worry about ordering the operators.
There are different conventions when one goes from a classical to quantum Hamiltonian. Here we chose the Weyl ordering because it appears to be a natural choice if we look at the metric in the Eddington-Finkelstein coordinates. The mixed term in the Hamiltonian ($\pi b$) arises due to the cross term in the metric and it is actually of the form ($\frac{\pi b + b \pi}{2}$). Therefore, Weyl prescription preserves the natural order of terms and treats $b$ and $\pi$ on equal footing. Thus, we write
\be
\pi b \rightarrow   \frac{(\hat{\pi}\hat{b}+ \hat{b}\hat{\pi})}{2} .
\ee
With this prescription, the standard Schrodinger equation
\be
H \psi = i\frac{\partial\psi}{\partial v}
\ee
is given as
\begin{equation}
   i\frac{\partial\psi}{\partial v}=  \frac{\lambda}{m} \left[ - \frac{1}{2} \frac{\partial^2 }{\partial b^2} - \frac{i y}{2}\left(1+ 2b\frac{\partial}{\partial b}\right) + \frac{1}{2} (y^2 +l m +\frac{k m}{\lambda}) b^2 \right] \psi .
   \label{partialSchrod}
\end{equation}
We now use an ansatz for the solution
\be
\psi(b,v) = \exp\big{[}-i\frac{yb^2}{2}\big{]} \phi (b,v) .
\ee
After little algebra we arrive at
\begin{equation}
   i\frac{\partial\phi}{\partial v}=  \frac{\lambda}{m} \left[ - \frac{1}{2} \frac{\partial^2 }{\partial b^2} + \frac{1}{2} (l m +\frac{k m}{\lambda}) b^2 \right] \phi
\end{equation}
As we can see, our ansatz significantly simplifies the equation that we have to solve.  We now define a new time parameter
\be
  \eta=\int dv'\sqrt{{R_v}^2 - 2 R_v +B},
\ee
which removes the time dependence from the coefficient in front of the kinetic term and reduces the Schrodinger equation to
\be
   i\frac{\partial\phi}{\partial \eta}=  \left[ - \frac{1}{2m} \frac{\partial^2 }{\partial b^2} + \frac{m}{2}\left (\frac{l}{m} +\frac{k/ m}{\lambda}\right) b^2 \right] \phi .
\ee
This equation can be written in the standard form as
\be
  \left[-\frac{1}{2m}\frac{\partial^2}{\partial b^2}+\frac{m}{2}\omega^2(\eta) \right]\phi(b,\eta)=i\frac{\partial\phi(b,\eta)}{\partial\eta} ,
  \label{ordinarySchrod}
\ee
where
\begin{equation} \label{omega}
  \omega^2(\eta) = \frac{l}{m} + \frac{{\omega_0}^2}{\sqrt{{R_v}^2 - 2 R_v +B}} .
\end{equation}
Here, we defined $\omega_0^2\equiv k/m$ which is just the initial frequency of the mode at the time it was created. Since background is time dependent, $\omega$ will evolve as time progresses. The time dependence is encoded inside $\lambda = \sqrt{{R_v}^2 - 2R_v +B}$. The term $l/m$ is a constant shift in the frequency which depends on the eigenvalues of the matrices $A$ and $D$. The solution of Eq.(\ref{ordinarySchrod}) is
\be
  \phi(b,\eta)=e^{i\alpha(\eta)}\left(\frac{m}{\pi\theta^2}\right)^{1/4}\exp\left[\frac{im}{2}\left(\frac{\theta_{\eta}}{\theta}+\frac{i}{\theta^2}\right)b^2\right]
  \label{phi} ,
\ee
where $\theta(\eta)$ is the solution of the ordinary differential equation given by
 \be
 \theta_{\eta\eta}+\omega^2(\eta)\theta=\frac{1}{\theta^3}.
 \label{thetaeq}
\ee
 Therefore, we converted a partial differential equation into a ordinary differential equation which are much easier to handle. The initial conditions for $\theta$ are taken at some large value of $\eta$ (which corresponds to a large value of $R$) denoted by $\eta_i$, so that
\be
  \theta(\eta_i)=\frac{1}{\sqrt{\omega(\eta_i)}}, \hspace{2mm} \theta_{\eta}(\eta_i)=0,
\ee
where the subscript $v$ stands for the derivative w.r.t $v$. The phase factor in Eq.~(\ref{phi}) is
\be
  \alpha(\eta)=-\frac{1}{2}\int^{\eta}\frac{d\eta'}{\theta^2(\eta')}.
\ee
If we transform Eq.(\ref{thetaeq}) to coordinate $v$, it becomes
\be
\theta_{vv} - \frac{\lambda_v}{\lambda}\theta_v +\lambda^{2} \omega^2(v) \theta  -\frac{ \lambda^{2}}{ {\theta}^3} = 0 .
\label{thetaeq_v}
\ee
In Eq. (\ref{thetaeq_v}), $\omega^2 (v)$ is the frequency measured by an observer with $\eta$ coordinate, just written in terms of $v$ by using appropriate transformation from $\eta$ to $v$. The proper frequency measured by an infalling observer is actually given as
\bea
\Omega(v) =& \frac{d \eta}{dv} \omega (\eta)
                  =& \lambda \omega (v)\\
                  =& \sqrt{\frac{ l {\lambda}^2}{m}+ {{\omega}_0}^2 \lambda} \label{Omega} .
\eea
 Here both terms are time dependent, so for $\omega_0 =0$, $\Omega$ can be non zero.

The total wavefunction which is a solution to the Schrodinger equation can now be written as
\be
\psi(b, v) = Exp\left[ -\frac{iy}{2} b^2\right]\phi(b,v) .
\ee
We can expand this wavefunction in terms of simple harmonic oscillator basis, which provide a natural basis for particle excitations, as
\be
  \psi(b,v)=\sum_nc_n(v)\varphi_n(b),
  \label{wavefnexpan}
\ee
with the expansion coefficients
\be
  c_n=\int db\varphi_n^*(b)\psi(b,v) .
  \label{coeff}
\ee
This definition is usual in the context of the functional Schrodinger formalism (see e.g. \cite{Vachaspati:2006ki,Vachaspati:2007hr,Greenwood:2010mr,Halstead:2011gs,Greenwood:2010uy,Greenwood:2009pd,Greenwood:2010sx}).
For even values of $n$ (see the appendix), $c_n$ are given by
\begin{equation}
  c_n(v)=\frac{(-1)^{n/2}e^{i\alpha}}{(\Omega \lambda^{-1}\theta^2)^{1/4}}\sqrt{\frac{2}{P}}\left(1-\frac{2}{P}\right)^{n/2}\frac{(n-1)!!}{\sqrt{n!}}
  \label{cneq}
\end{equation}
where  $P$ is
\begin{equation}
  P=1-\frac{i \lambda}{\Omega}\left(\frac{\theta_v}{\theta \lambda}+\frac{i}{\theta^2}\right) + \frac{iy \lambda}{ m \Omega} .
\end{equation}
All $c_n$ for odd values of $n$ vanish. The occupation number of the excited particles as a function of frequency $\Omega$ by the time $v<v_f$ (where $v_f$ is the final time when we count the excited modes), is given by
\be
  N(v,\Omega)=\frac{\Omega\theta^2}{4 \lambda}
  \left[\left(1-\frac{\lambda}{\Omega\theta^2}\right)^2+\left(\frac{\theta_{v}}{\Omega\theta} - \frac{y \lambda}{ m \Omega}\right)^2\right] .
\ee

\section{Radiation spectrum}

\begin{figure}[htpb]
\begin{center}
  \includegraphics[height=0.30\textwidth,angle=0]{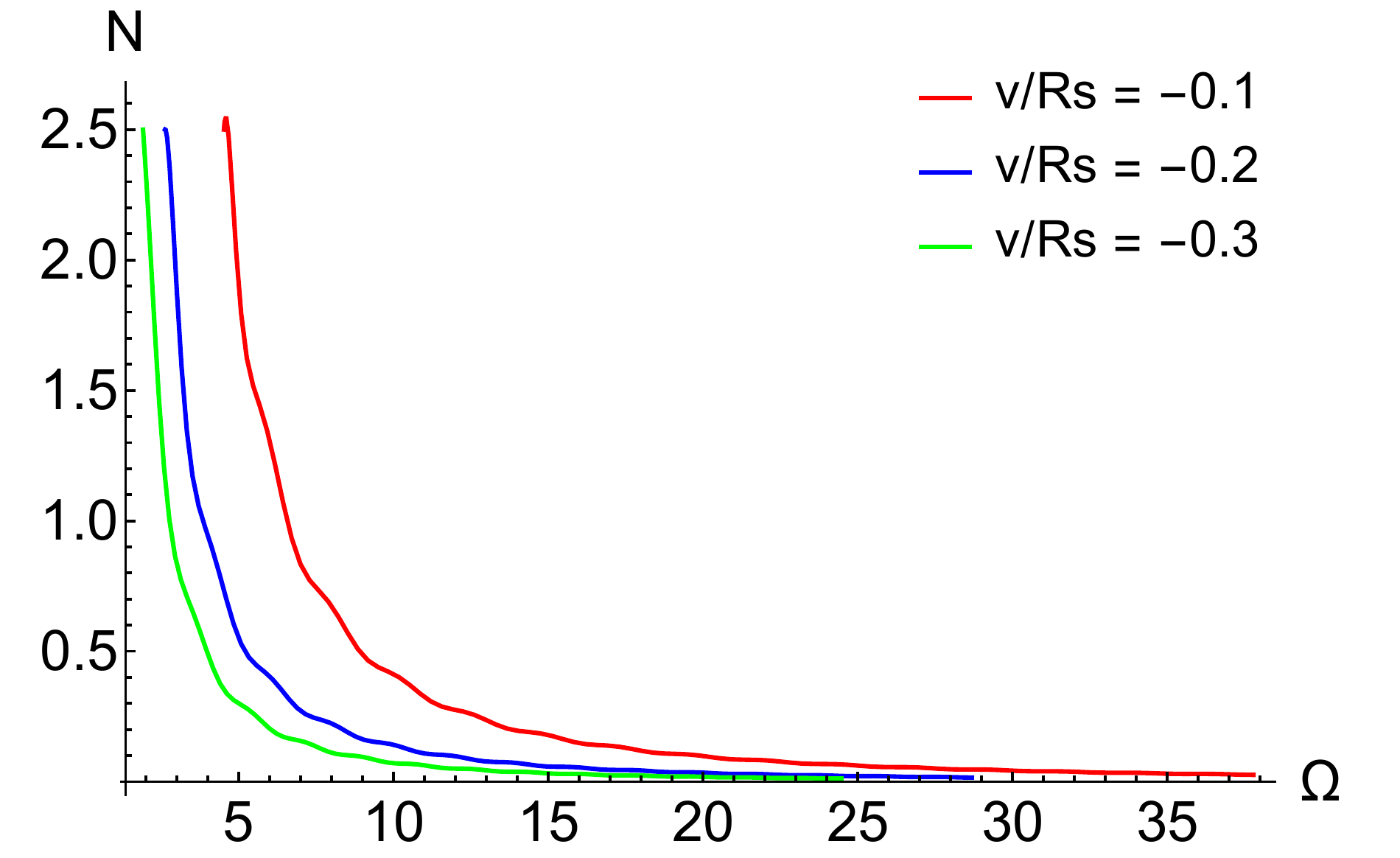}
\caption{ N(v, $\Omega$) is plotted as a function of $\Omega$ at several different times. We set $y = 0.01$, $l=0.00001$, $R_s = 100$ and $\sigma =3*10^{-6}$. Shell crosses its event horizon at $v/R_s =0$ and approaches singularity at $v/R_s =  5.68*10^{-6}$.  The particle spectrum is following approximately Planckian distribution with some fluctuations. Low frequencies are not excited due to finite value of $l$. As $v/R_s$ increases minimum excitable frequency increases.}
\label{Nvsomega}
\end{center}
\end{figure}

\begin{figure}[htpb]
\begin{center}
  \includegraphics[height=0.30\textwidth,angle=0]{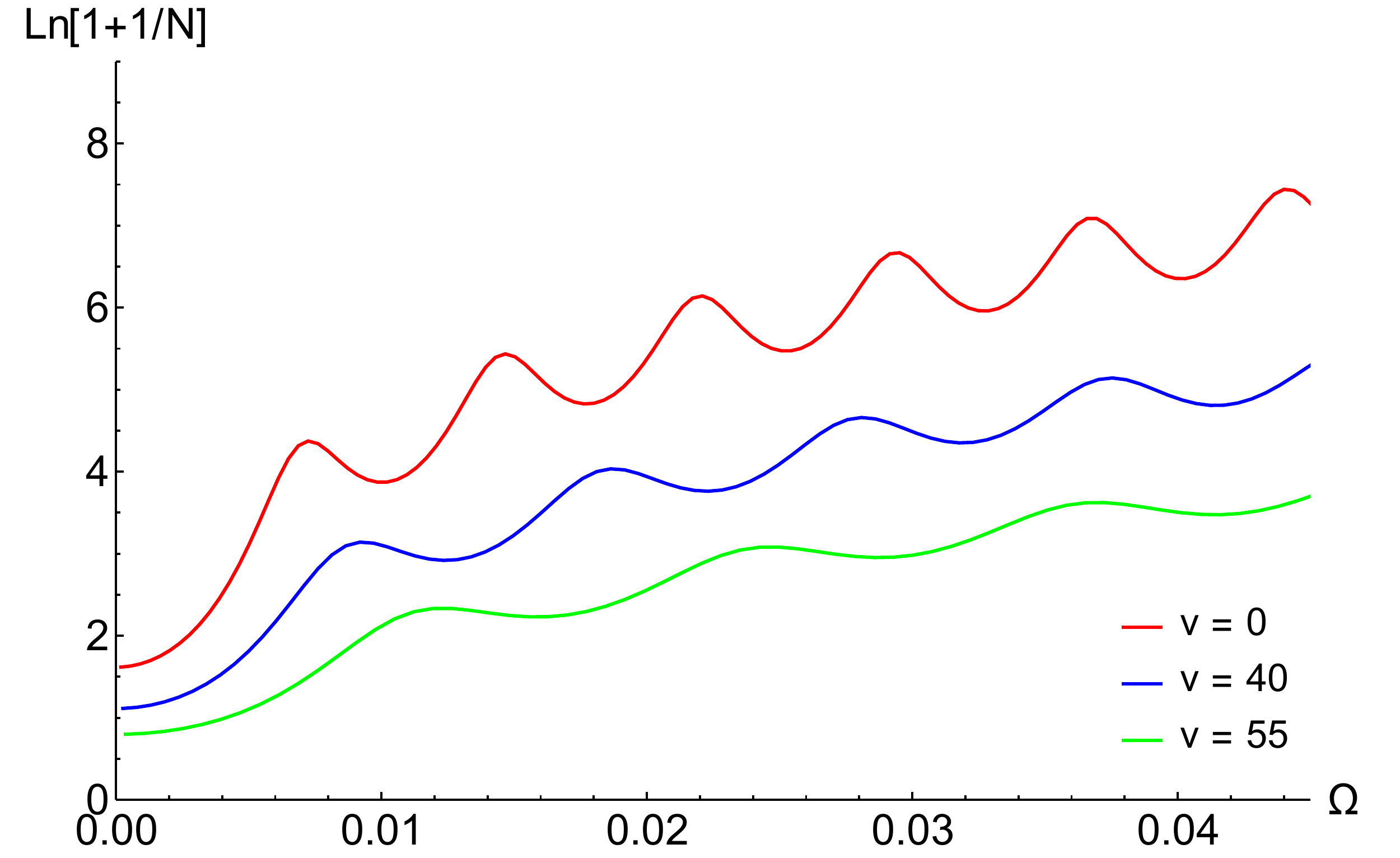}
\caption{ In this figure, we plotted $\ln(1+1/N)$ as a function of $\Omega$ at various time. For simplicity we put $l= y =0$. We chose $R_s = 1000$ and $ \sigma =3 *10^{-5}$. The slope of $\ln(1+1/N)$ vs $\Omega$ is inversely related to temperature of the emitted radiation. The curve can be approximated linear for $\Omega <0.01$. We fitted the curve in this range and got $T \sim0.002 $, $T \sim0.003 $, $T \sim0.006 $ at $v = 0$  (at the moment when the shell is crossing its own Schwarzschild radius), $v =40$ and $v =55$ respectively. As expected, temperature is increasing as $v$ is increasing (i.e. time is progressing).}
\label{LnNvsomega}
\end{center}
\end{figure}

As stated earlier, Eq.~(\ref{rvdynamics}) can be used to find the classical evolution of the collapsing shell in infalling observer coordinates. $R(v)$ can be found by numerically solving differential equation given as
\be
R_v(v) =  B{\left(1-\sqrt{1+ \frac{B}{{\left( \frac{M}{4 \pi G R(v)^2}+ 2 \pi \sigma G R(v)\right)}^2-1}}\right)}^{-1}
\label{rvexact}
\ee
which then enters the expression for $\lambda$ in Eq.~(\ref{lambda}) and all the subsequent equations. We set up the initial condition of the system so that shell crosses its own Schwarzschild radius at $v=0$. It is important that the $R_{max}< \frac{1}{4\pi \sigma G}$ otherwise shell will be within its own Schwarzschild radius at the beginning. We set eigenvalue $m$ and Newton's constant $G$ equal to unity for the purpose of numerical calculations. In Fig.(\ref{Nvsomega}), the occupation number of excited modes $N(v/R_s,\Omega)$ is plotted as a function of $\Omega$ at different times, $v/R_s$, for the eigenvalues $y = 0.01$, $l = 0.00001$. We see that the spectrum follows near Planckian distribution (but it is not exactly Planckian).  This fact will allow us to fit an approximate temperature, $T$, of radiation since the Planckian distribution goes as $1/(e^{\Omega/T}-1)$. It is also noticeable that all frequencies are not excited due to finite value of the eigenvalue $l$. In Eq.~(\ref{Omega}), $\omega_0 =0$ gives $\Omega_{min} =\lambda(v) \sqrt{l/m}$, which means there is a lower bound on frequency, $\Omega_{min}$, below which the modes are not excited. Since $\lambda$ is a monotonically increasing function of $v$, $\Omega_{min}$ increases as the shell approaches singularity. \\
\\
In Fig.~(\ref{LnNvsomega}), we plot  $\ln(1+1/N)$ as a function of $\Omega$ which will gives us an estimate of the temperature of the emitted radiation.  The slope of this graph  is inversely proportional to temperature of the radiation. Since the distribution in Fig.~\ref{Nvsomega} is not thermal (Planckian), this analysis will only give rough estimate. We see that  slope of the curve is decreasing as $v$ increases. This is expected because as shell approaches the singularity, more  higher energy states get excited and temperature increases. The initial stage of the collapse, when the shell is far from its own Schwarzschild radius, exhibits highly non thermal features to draw any conclusion about its temperature. Though distribution never seems to become thermal in the whole frequency range, one can still fit the temperature for wavelengths which are comparable to the Schwarzschild radius when the collapsing shell is close to it. Using linear fit for $\Omega < 0.01$, we find out that $T_B \sim 0.002 $ at $v =0$, for   $R_s = 1000$ and $ \sigma =3 *10^{-5}$. We found that the temperature decreases with increasing mass and surface density of shell.  This is consistent with standard result that the black hole temperature is inversely proportional to its mass. We also find out that temperature increases monotonically with eigenvalue $y$. This trend can be seen in other graphs too.
\\

 While choosing values for $\sigma$ in order to generate the numerical plots, one has to be careful not to choose the values that yield a black hole to begin with. A large constant $\sigma$ together with large initial radius $R$ give an object which is already with its own Schwarzschild radius. The condition to avoid this case is $\max(R) < (4 \pi G \sigma)^{-1}$.

\section{Density matrix and information content}

Since our formalism gives us the full wavefunction of the excited radiation, we can now study the information content in it. This can be done by constructing the density matrix($\hat{\rho}$) for the system. Since we already know the $c_n$ (see Eq.(\ref{cneq})), individual elements of the density matrix can be evaluated numerically. A density matrix of a quantum system is defined as
\be
\hat{\rho} =  \ket{\psi}\bra{\psi} .
\ee
using Eq.(\ref{wavefnexpan}), we can re write density matrix as
\be
\hat{\rho}  = \sum_{m,n} c_{mn}\ket{ \phi_m}\bra{ \phi_n}
\ee
where $c_{mn} \equiv c_m*c_n$. The time evolution of the system is encoded in coefficients $c_m, c_m$. Only even terms in density matrix will survive due to the vanishing $c_n$ for odd values of n. The structure of matrix will be

\[
M=
  \begin{bmatrix}
    c_{00} & 0 & c_{02} & 0 & .. \\
      0  &  0&0&0&..\\
    c_{20} & 0 & c_{22} & 0 & ..\\
    0 & 0&0&0&..\\
    . & ..&..&..&..\\
    . & ..&..&..&..\\
  \end{bmatrix}
\]

\begin{figure}[htpb]
\begin{center}
  \includegraphics[height=0.30\textwidth,angle=0]{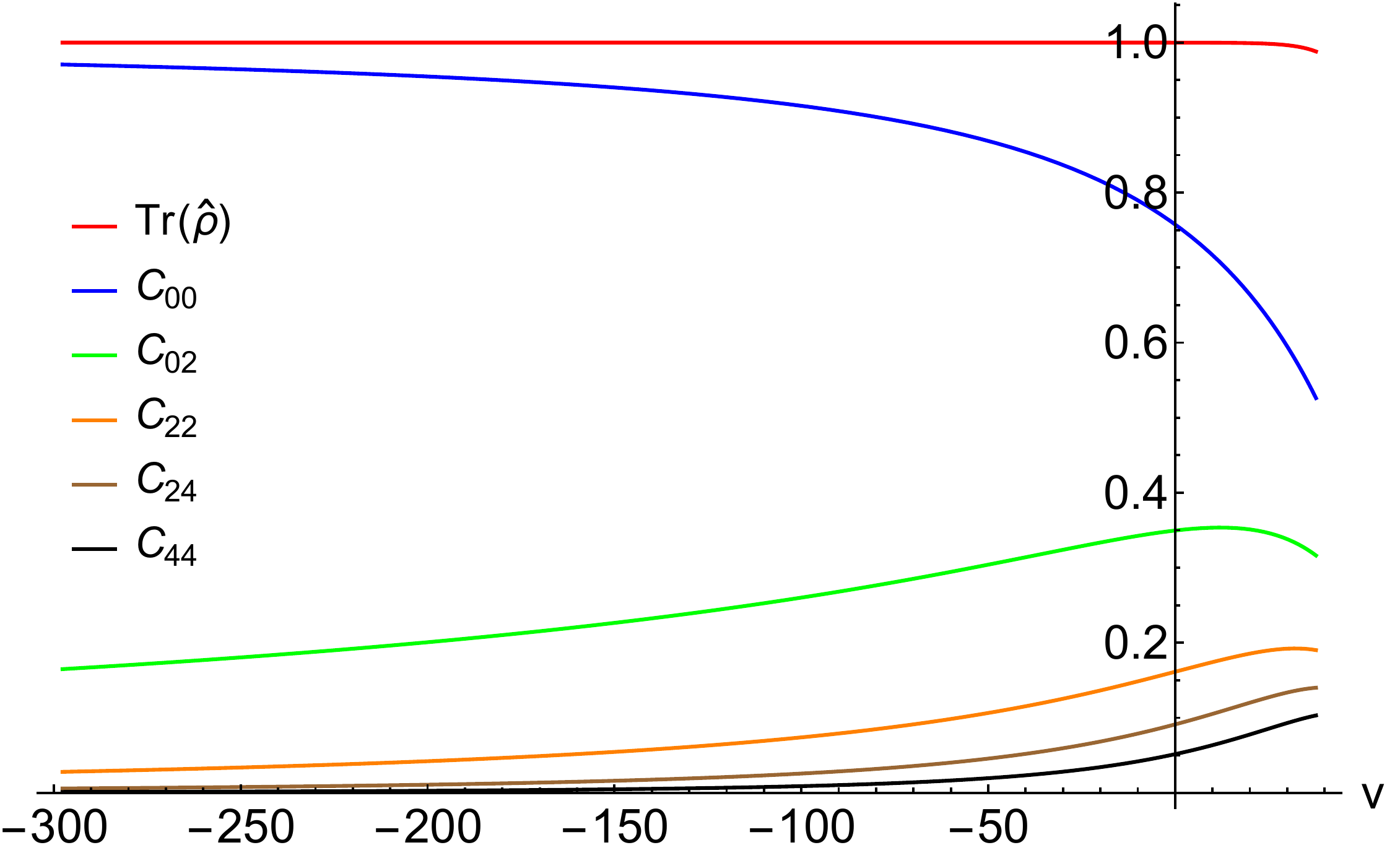}
\caption{We plot various density matrix elements $c_{mn}$ as a function of time, $v$, at $\Omega = 10$  almost to the time when the singularity is formed. We set  $ y = 5$, $ l =1$, $\sigma =3*10^{-5}$ and $R_s = 1000$. We see that $Tr(\hat{\rho})$ remains one at all times. Initially $c_{00}$ is one (vacuum) and other terms are zero, but as time increases higher terms increase.  the off diagonal $c_{mn}$ terms which represent correlations between the emitted particles are of the same order of magnitude as diagonal terms which represent particles.}
\label{cmnvsv}
\end{center}
\end{figure}

\begin{figure}[htpb]
\begin{center}
  \includegraphics[height=0.30\textwidth,angle=0]{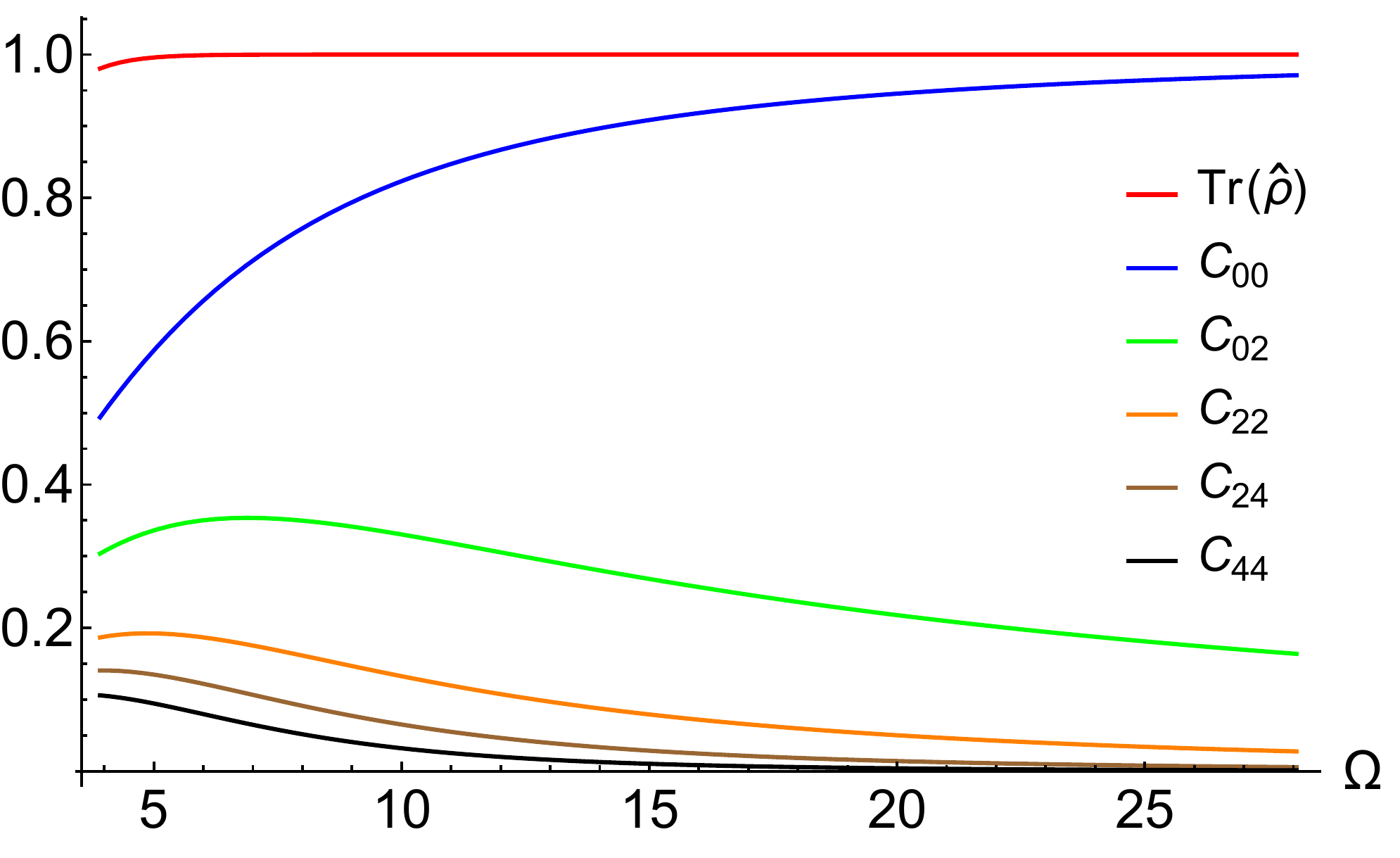}
\caption{We plot the density matrix elements $c_{mn}$ as a function of $\Omega$ at $v =0$. We fix $y = 4$, $ l = 1$, $\sigma =3*10^{-5}$ and $R_s = 1000$. $c_{00}$ increases with $\Omega$ while all the other $c_{mn}$ decrease with $\Omega$. This means that modes at higher frequencies at some fixed time prefer to stay in their vacuum. $Tr(\hat{\rho})$ remains unity at all $\Omega$.}
\label{cmnvsomega}
\end{center}
\end{figure}

If a system is in a pure state, then $Tr(\hat{\rho}^2)=1$. A state is mixed if $Tr(\hat{\rho}^2)<1$, with extreme case of a maximally mixed state with $Tr(\hat{\rho}^2) = 0$. Such system contains no information in it.
In Fig.(\ref{cmnvsv}) we  plotted $Tr(\hat{\rho})$ and various $c_{mn}$ as a function of time at fixed frequency($\Omega$). We extend the plot almost to the time when the singularity is formed, which happens at $v=66.54$. At that point all the calculations break down. We can see that $Tr(\hat{\rho})$ remains one at all times, including the time after the Schwarzschild radius is passed at $v =0$. Initially the coefficient $c_{00}$ is one, while all higher excited modes are zero meaning that the system starts from vacuum. As $v$ increases higher states are excited due to time dependent spacetime, so all of the higher terms $c_{mn}$ increase while $c_{00}$ decreases. Most importantly, the off diagonal $c_{mn}$ terms which represent correlations between the emitted particles are of the same order of magnitude as diagonal terms which represent particles. This is important because the correlations are often ignored under the assumption that they are negligible. However we will see that if the off-
diagonal terms are ignored then one would erroneously conclude that $Tr(\hat{\rho}^2)<1$ and information is lost.
\\
To get some more information, in Fig.(\ref{cmnvsomega}) we plot $Tr(\hat{\rho})$ and various $c_{mn}$ as a function of infalling observer's frequency $\Omega$ at a fixed time $v=0$. We find that $Tr(\hat{\rho})$ remains one for all frequencies as it should. We also see that $c_{00}$ increases as frequency increases but all the other elements decrease with frequency. This means that modes at higher frequencies are more difficult to excite and they prefer to stay in their vacuum. Again there is a lower bound on allowed frequency range due to finite value of $l$.

\begin{figure}[htpb]
\begin{center}
  \includegraphics[height=0.30\textwidth,angle=0]{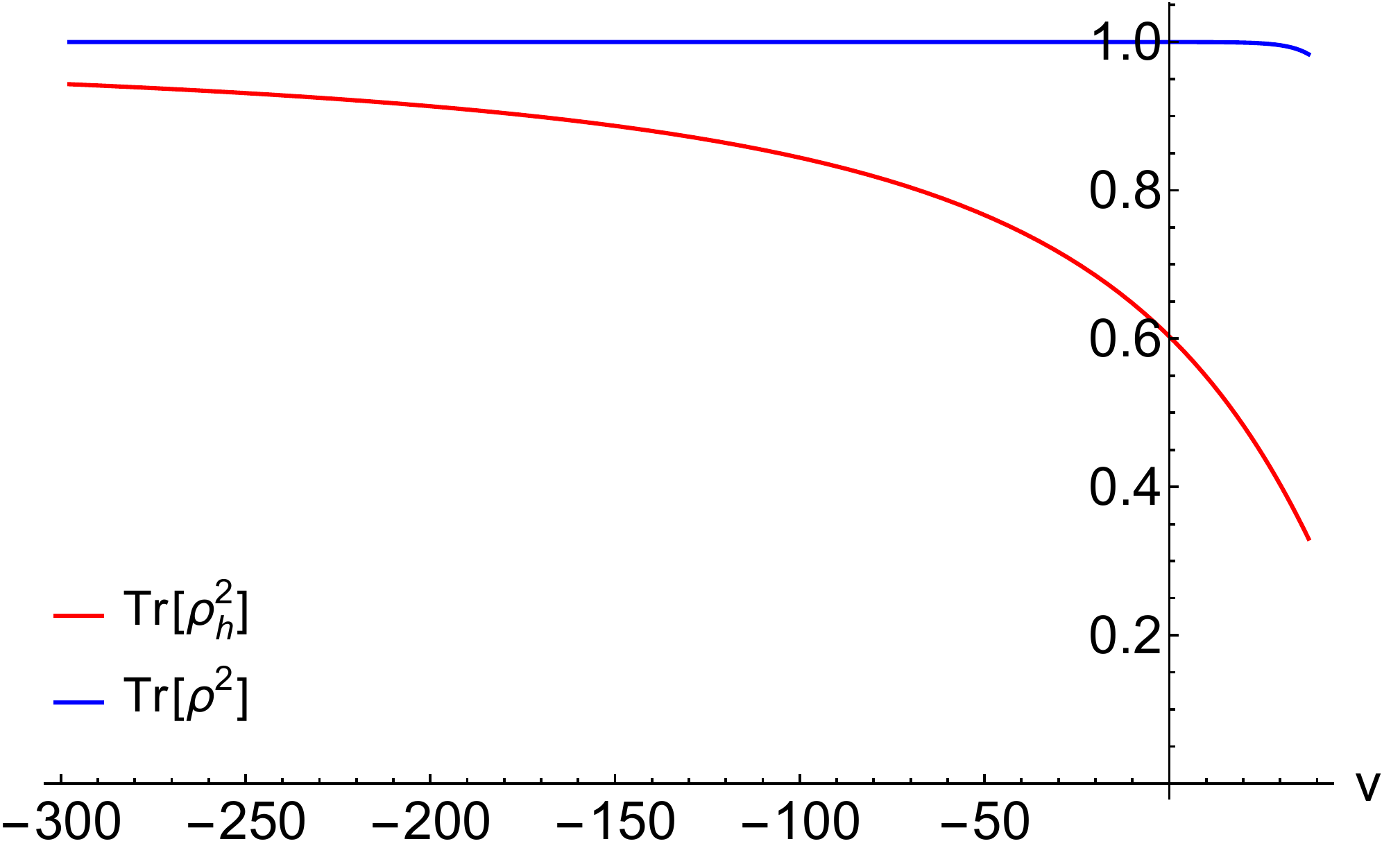}
\caption{Here we plot the trace of the squared density matrix which includes all the elements, Tr(${\rho^2}$), and the trace of the squared density matrix which includes only the diagonal terms (Hawking quanta), Tr(${\rho_h^2}$), at $\Omega =10$ as a function of time. Other parameters chosen to be $y = 5$, $ l= 1$, $\sigma =3*10^{-5}$ and $R_s = 1000$.  Tr($\rho_h^2$) starts from unity, but then decreases as time progresses indicating information loss. On the other hand  Tr(${\rho^2}$) remains unity all the time, which means state is pure at all times due to the significant contribution from off diagonal elements.}
\label{trvsv}
\end{center}
\end{figure}

Now we want to contrast the total density matrix $\rho$, which contains the complete information about the system (both diagonal and off-diagonal terms), with the density matrix  $\rho_h$, which contains only the diagonal terms, i.e. Hawking quanta.
In Fig.~(\ref{trvsv}) we simultaneously plot Tr($\rho^2$) and Tr($\rho_h^2$) as a function of time at fixed $\Omega$. As we can see from the plot,  Tr(${\rho^2}_h$) starts at unity indicating that the quantum state is pure. As the time is progressing, Tr(${\rho^2}_h$) is decreasing indicating that the system is evolving from the pure to a mixed state. However  Tr(${\rho^2}$) is unity at all times, so if off-diagonal terms are also taken into account the system remains pure. This means that correlations between the emitted quanta are crucial in preserving unitarity of the evolution. Obviously, the state remains pure even after the shell crosses its own Schwarzschild radius at $v=0$. For technical reasons we could not extend our graph all the way to the singularity because higher and higher modes get rapidly excited, while we can take only finite number of terms  in our density matrix.

\begin{figure}[htpb]
\begin{center}
  \includegraphics[height=0.30\textwidth,angle=0]{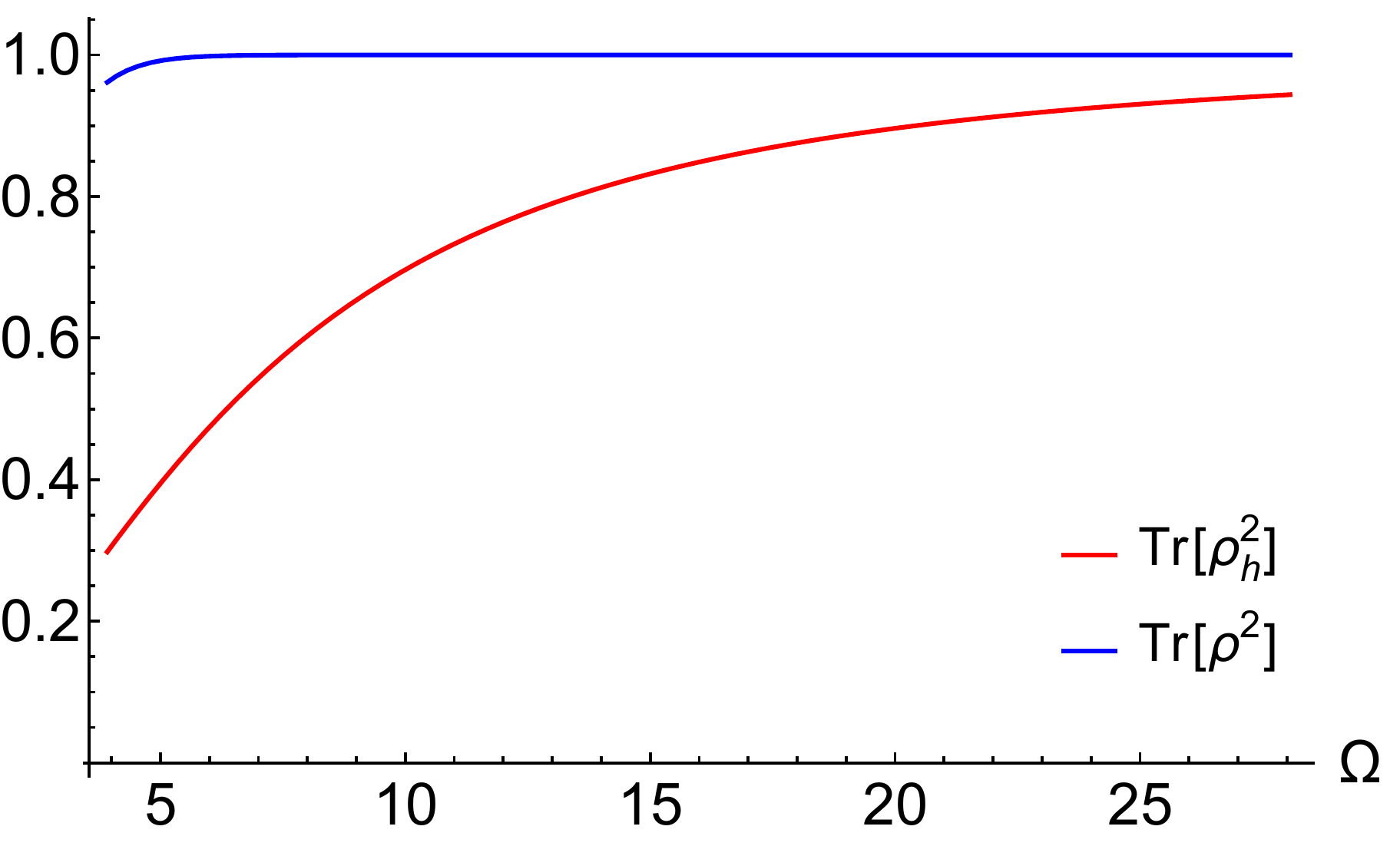}
\caption{ Tr($\rho^2$) and Tr($\rho_h^2$) are plotted at fixed $v = 0$ as a function of $\Omega$.  Relevant constants are chosen as $ y = 4$ and $ l= 1$, $\sigma =3*10^{-5}$ and $R_s = 1000$. Tr(${\rho^2}_h$) is close to unity only at higher frequencies where $c_{00}$ is dominant and higher terms are not excited. However   Tr($\rho^2$) is unity at all frequencies.}
\label{Trt}
\end{center}
\end{figure}

\begin{figure}[htpb]
\begin{center}
  \includegraphics[height=0.30\textwidth,angle=0]{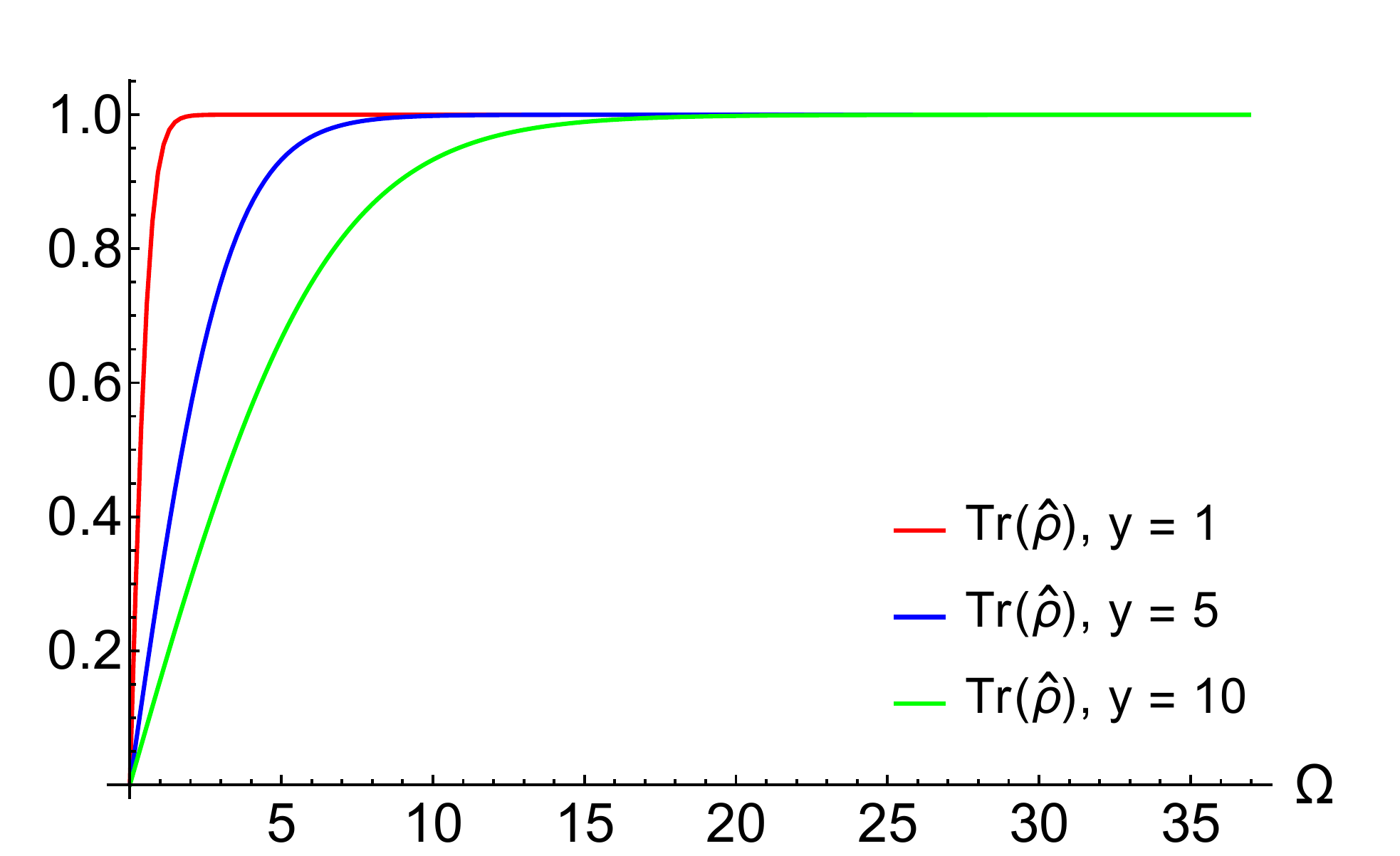}
\caption{Here we plot Tr($\rho$) for different value of the eigenvalue $y$ keeping $l =0$, $R_s =1000$ and $\sigma =3* 10^{-5}$ at $v =-0.1$. We can see that as $y$ increases, Tr($\rho$)$<1$ at low frequencies, which indicates that one has to increase the number of calculated terms $c_{mn}$ in order to bring Tr($\rho$) back to unity. Hence higher order terms are more prominent at higher value of $y$. }
\label{trace_yeffect}
\end{center}
\end{figure}

In Fig.~\ref{Trt}, we plot  Tr($\rho^2$) and Tr(${\rho^2}_h$) as a function of $\Omega$ at the fixed time $v=0$. Again  Tr($\rho^2$) remains unity for all frequencies indicating that the total state is pure.  Tr(${\rho^2}_h$) is close to one at higher frequencies which means that the correlations at higher frequencies are less important.
This happens because at high frequencies the term $c_{00}$ is dominant and higher terms are not excited.
\\
From Eq.(\ref{cneq}), we can see that $c_n$  is depends on the eigenvalue of $y$. To study this dependence,  In Fig.~(\ref{trace_yeffect}) we plot  Tr($\rho$) for different values of $y$ while keeping same value of $v =0$ and $l$. We see that, for large values of $y$, the plot of Tr($\rho$) which can technically incorporate only a finite number of terms departs from unity at low frequencies. In that regime the calculations cannot be trusted and one has to increase the number of calculated terms $c_{mn}$.
\\

\section{Conclusions}
\label{conclusions}
In this paper we studied radiation emitted by a collapsing shell in Eddington-Finkelstein spacetime foliation which corresponds to an infalling observer.
During the gravitational collapse, the spacetime metric is time-dependent which leads to particle production even in Eddington-Finkelstein frame.
Naively, equivalence principle might imply that an infalling observer should not register any radiation since he is not an accelerated observer. This statement is certainly true for a non-accelerated observer in Minkowski spacetime. However, in a curved spacetime, an infalling observer is in a effectively flat spacetime only in an infinitesimal neighbourhood of a point, so such an observer might not observe all the wavelengths of radiation but he would certainly observe some of them.
We performed explicit calculations in the framework of the functional Schrodinger formalism to find the analytic form of the spectrum of excited modes. We did not use any approximations, so our results are exact, though the final plots are calculated numerically. We find that the spectrum is thermal only in a limited range of frequencies. By performing the best fit through the approximately thermal part of the spectrum, we find that the temperature of radiation increases as the collapse progresses. When the shell is near its own Schwarzschild radius, the temperature is finite and close to the Hawking temperature. As the shell is collapsing to a point and approaching singularity, the temperature grows without limits (this conclusion will of course be evaded if quantum effects are able to remove the classical singularity from the center  \cite{FV,Saini:2014qpa,Greenwood:2008ht,Wang:2009ay,Bogojevic:1998ma,Kunstatter:2015vxa,Taves:2014laa,Bojowald:2005ah,Easson:2002tg,Culetu:2015cna,Bronnikov:2006fu,Guendelman:2009vz}).

Our formalism gives the full wavefunction of the excited radiation, and we studied the information content in it by constructing the density matrix for the system.
We showed that it is not enough to keep only the diagonal terms (particles) in the density matrix. This would lead to an apparent information loss, since a trace of the square of such a density matrix would depart from unity during the collapse. In contrast, if we include the off-diagonal terms which encode correlations between the diagonal modes then the trace of the square of the density matrix remains unity at all times and all frequencies, indicating unitary evolution from the beginning till the end of the collapse. Once the singularity is encountered, the whole formalism breaks down, at which point the infalling observer presumably does not exist anymore. At that point the equations of motions become singular, so we cannot extend the analysis beyond this point.

While the standard infirmation loss paradox is formulated for an asymptotic observer, it was still instructive to calculate the density matrix for this case too. If nothing else, we learned that in this case too the diagonal terms in the density matrix are not sufficient, and most of the infirmation is actually hidden in the correlation between the diagonal terms \cite{Saini:2015dea}, which agrees well with Page's argument made for a thermodynamical system in the absence of gravity \cite{Page:1993df}.  Since the infalling observer ultimately ends up at the singularity, one might try to exclude part of the outgoing radiation that such an observer might not be able detect. This would lead to a incomplete density matrix and unitarity would seem to be broken. 

\begin{acknowledgments}
This work was partially supported by NSF, grant number PHY-1417317.
\end{acknowledgments}

\appendix

\section{Number of particles produced as a function of time}

We use the simple harmonic oscillator basis states with frequency $\Omega$ with respect to the infalling observer. The mass of the modes ($M$) with respect to infalling observer will also be changed accordingly and it is related to mass($m$) in $\eta$ coordinates as
\be
M = \frac{m}{\lambda}
\ee
The wavefunction of harmonic oscillator states can be written as

\begin{equation}
  \varphi(b)=\left(\frac{M\Omega}{\pi}\right)^{1/4}\frac{e^{-M \Omega b^2/2}}{\sqrt{2^nn!}}H_n(\sqrt{M\Omega}b)
\end{equation}
where $H_n$ are the Hermite polynomials. Then Eq.~(\ref{wavefnexpan}) together with Eq.~(\ref{coeff}) gives
\begin{align}
  c_n=&\left(\frac{\lambda}{\Omega\pi^2\rho^2}\right)^{1/4}\frac{e^{i\alpha}}{\sqrt{2^nn!}}\int d\zeta e^{-P\zeta^2/2}H_n(\zeta)\nonumber\\
           \equiv&\left(\frac{\lambda}{\Omega\pi^2\rho^2}\right)^{1/4}\frac{e^{i\alpha}}{\sqrt{2^nn!}}I_n
\end{align}
where
\begin{equation}
  P=1-\frac{i \lambda}{ \Omega}\left(\frac{\rho_v}{\lambda \rho}+\frac{i}{\rho^2}-\frac{y}{m}\right).
\end{equation}

To find $I_n$ consider the corresponding integral over the generating function for the Hermite polynomials
\begin{align}
  J(z)&=\int d\zeta e^{-P\zeta^2/2}e^{-z^2+2z\zeta}\nonumber\\
         &=\sqrt{\frac{2\pi}{P}}e^{-z^2(1-2/P)}
\end{align}
Since
\begin{equation}
  e^{-z^2+2z\zeta}=\sum_{n=0}^{\infty}\frac{z^n}{n!}H_n(\zeta)
\end{equation}
\begin{equation}
  \int d\zeta e^{-P\zeta^2/2}H_n(\zeta)=\frac{d^n}{dz^n}J(z)\Big{|}_{z=0}
\end{equation}
Therefore
\begin{equation}
  I_n=\sqrt{\frac{2\pi}{P}}\left(1-\frac{2}{P}\right)^{n/2}H_n(0).
\end{equation}
Since
\begin{equation}
  H_n(0)=(-1)^{n/2}\sqrt{2^nn!}\frac{(n-1)!!}{\sqrt{n!}}
\end{equation}
and $H_n(0)=0$ for odd $n$, we find the coefficient $c_n$ for even values of $n$,
\begin{equation}
  c_n=\frac{(-1)^{n/2}e^{i\alpha}}{({\lambda}^{-1}{\Omega}\rho^2)^{1/4}}\sqrt{\frac{2}{P}}\left(1-\frac{2}{P}\right)^{n/2}\frac{(n-1)!!}{\sqrt{n!}}.
\end{equation}
For odd $n$, $c_n=0$.

\section{Trace calculations}

In this part we will prove that the trace of the system will add up to unity. Lets take
\begin{equation}
  \xi=\left|1-\frac{2}{P}\right|.
\end{equation}
Now,

\begin{align}
  Tr(\rho)&=\sum_{n=even}\left|c_n\right|^2\nonumber\\
                                  &=\frac{2}{\sqrt{\Omega\rho^2 \lambda^{-1}}|P|}\sum_{n=even}\frac{(n-1)!!}{n!!}\xi^n\nonumber\\
                                  &=\frac{2}{\sqrt{\Omega\rho^2 \lambda^{-1}}|P|}\frac{1}{\sqrt{1-\xi^2}}\nonumber\\
                                  &=\frac{2}{\sqrt{\Omega\rho^2 \lambda^{-1}}|P|}\frac{1}{\sqrt{1-{\left|  1- \frac{2}{P}  \right|}^2}}
\end{align}
Inserting the expressions for $P$ and little algebra gives  $Tr(\rho) = 1$.

\end{document}